\documentclass[letterpaper,twocolumn,nofootinbib,showkeys,superscriptaddress,unsortedaddress,amsfonts,amssymb,amsmath]{revtex4}
\usepackage{mysty}
\usepackage{multirow}
\usepackage{times,mathptmx}
\usepackage[utf8]{inputenc}
\usepackage[T1]{fontenc}
\usepackage{hyperref}
\usepackage{amsmath,amssymb,amsfonts}
\usepackage[pdftex,dvipsnames]{xcolor}
\usepackage{titlesec}
\usepackage{graphicx}
 
\hypersetup{
    unicode=true,          
    pdftoolbar=true,        
    pdfmenubar=true,        
    pdfstartview={FitH},    
    pdftitle={Structural basis for the shielding function of the dynamic trypanosome variant surface glycoprotein coat}, 
    pdfauthor={Mislav Cvitković}, 
    pdfsubject={Structural basis for the shielding function of the dynamic trypanosome variant surface glycoprotein coat}, 
    pdfcreator={Mislav Cvitković},   
    pdfproducer={Mislav Cvitković}, 
    pdfkeywords={Membrane biophysics, Membrane trafficking, Parasite immune evasion, SAXS}, 
    colorlinks=true,       
    linkcolor=blue,          
    citecolor=purple,        
    filecolor=magenta,      
    urlcolor=magenta           
}

\makeatletter
\renewcommand*{\fnum@figure}{{\sffamily\footnotesize\bfseries \figurename~\thefigure{}.}}
\makeatother

\newcommand*{\doi}[1]{Doi:\ \href{http://dx.doi.org/#1}{#1}}

\renewcommand*{\url}[1]{Link:\ \href{#1}{#1}}
\newcommand{\musq}{$\mu$m${}^2$}
\newcommand{\degc}{$\,{}^\circ$C}

\begin{document}

\title{Structural basis for the shielding function of the dynamic trypanosome variant surface glycoprotein coat}

\author{Thomas Bartossek}
\affiliation{Department of Cell and Developmental Biology, Theodor-Boveri-Institute, Biocenter, University of Würzburg, 97074, Würzburg, Germany}

\author{Nicola G.\ Jones}
\thanks{T.\ Bartossek and N.\ G.\ Jones contributed equally to this work}
\email[E-mail: ]{nicola.jones@uni-wuerzburg.de}
\affiliation{Department of Cell and Developmental Biology, Theodor-Boveri-Institute, Biocenter, University of Würzburg, 97074, Würzburg, Germany}

\author{Christin Sch\"afer}
\affiliation{Rudolf Virchow Center for Experimental Biomedicine, Institute for Structural Biology, University of Würzburg, 97080, Würzburg, Germany}

\author{Mislav Cvitkovi\'c}
\affiliation{PULS Group, Institut für Theoretische Physik and the Excellence Cluster: Engineering of Advanced Materials, Friedrich-Alexander-Universität Erlangen-Nürnberg, 91052, Erlangen, Germany}
\affiliation{Group for Computational Life Sciences, Division of Physical Chemistry, Ruđer Bošković Institute, 10000, Zagreb, Croatia}

\author{Marius Glogger}
\affiliation{Department of Cell and Developmental Biology, Theodor-Boveri-Institute, Biocenter, University of Würzburg, 97074, Würzburg, Germany}

\author{Helen R.\ Mott}
\affiliation{Department of Biochemistry, University of Cambridge, Cambridge, CB2 1GA, UK}

\author{Jochen Kuper}
\affiliation{Rudolf Virchow Center for Experimental Biomedicine, Institute for Structural Biology, University of Würzburg, 97080, Würzburg, Germany}

\author{Martha Brennich}
\affiliation{European Synchrotron Radiation Facility, 71 Avenue des Martyrs, CS 40220, 38042, Grenoble, France}
\altaffiliation[Present address: ]{European Molecular Biology Laboratory, 71 Avenue des Martyrs, BP 181, 38042, Grenoble, France}

\author{Mark Carrington}
\affiliation{Department of Biochemistry, University of Cambridge, Cambridge, CB2 1GA, UK}

\author{Ana-Sun\v{c}ana Smith}
\affiliation{PULS Group, Institut für Theoretische Physik and the Excellence Cluster: Engineering of Advanced Materials, Friedrich-Alexander-Universität Erlangen-Nürnberg, 91052, Erlangen, Germany}
\affiliation{Group for Computational Life Sciences, Division of Physical Chemistry, Ruđer Bošković Institute, 10000, Zagreb, Croatia}

\author{Susanne Fenz}
\affiliation{Department of Cell and Developmental Biology, Theodor-Boveri-Institute, Biocenter, University of Würzburg, 97074, Würzburg, Germany}

\author{Caroline Kisker}
\affiliation{Rudolf Virchow Center for Experimental Biomedicine, Institute for Structural Biology, University of Würzburg, 97080, Würzburg, Germany}

\author{Markus Engstler}
\email[E-mail: ]{markus.engstler@biozentrum.uni-wuerzburg.de}
\affiliation{Department of Cell and Developmental Biology, Theodor-Boveri-Institute, Biocenter, University of Würzburg, 97074, Würzburg, Germany}

\begin{abstract}
\vspace{0.3cm}
The most prominent defence of the unicellular parasite \textit{Trypanosoma brucei} against the host immune system is a dense coat that comprises a variant surface glycoprotein (VSG). Despite the importance of the VSG family, no complete structure of a VSG has been reported. Making use of high-resolution structures of individual VSG domains, we employed small-angle X-ray scattering to elucidate the first two complete VSG structures. The resulting models imply that the linker regions confer great flexibility between domains, which suggests that VSGs can adopt two main conformations to respond to obstacles and changes of protein density, while maintaining a protective barrier at all times. Single-molecule diffusion measurements of VSG in supported lipid bilayers substantiate this possibility, as two freely diffusing populations could be detected. This translates into a highly flexible overall topology of the surface VSG coat, which displays both lateral movement in the plane of the membrane and variation in the overall thickness of the coat.
\end{abstract}

  \maketitle

%

\noindent The parasite  \textit{Trypanosoma brucei} thrives in the bloodstream and tissue spaces of its mammalian host, where it is prone to attack by the immune system. As a first defence against the immune response, the bloodstream form of trypanosomes is coated with a dense layer of variant surface glycoprotein (VSG). This layer constitutes at least 95\% of the cell-surface proteins\cite{Ziegelbauer1992,Gruenfelder2002} and impairs the identification of epitopes of invariant surface proteins by the immune system\cite{Cross1975,CardosodeAlmeida1983,Ziegelbauer1993,Sullivan2013}. Only a single VSG gene is transcribed at any given time from a specific genomic locus, the VSG expression site (ES). However, the VSG itself is highly immunogenic and, once detected by the immune system, antibodies specific for the expressed VSG isoform will facilitate a rapid clearance of the infection\cite{Macaskill1981,McLintock1993}. To establish a persistent infection, the parasite employs antigenic variation to switch the expression to an antigenically distinct VSG from a library that encodes several hundred VSGs\cite{Cross2014}. This continual switching forces the host immune system constantly to chase novel antigen types. The shielding function that VSG is thought to provide has been described only anecdotally and, owing to a lack of a complete VSG structural model, the mechanism remains unclear.

A further immune evasion strategy in \textit{T.~brucei} involves the rapid clearing of antibodies bound to VSGs through passive sorting by hydrodynamic drag forces\cite{Engstler2007}. These forces push the VSG--antibody complexes to the flagellar pocket, an invagination of the plasma membrane that spans approximately 5\% of the surface area of the pathogen. It is in this flagellar pocket that all endo- and exocytosis takes place and the complexes are rapidly internalized.

The VSGs analysed to date are homodimeric membrane proteins with a total size of 100--120 kDa that are connected to the outer leaflet of the plasma membrane by a glycosylphosphatidylinositol (GPI) anchor. VSG monomers consist of a small C-terminal domain (CTD) that connects to the larger N-terminal domain (NTD) via an unstructured region, Linker 1 (L1). This flexible linker is readily accessible to proteases, which is highlighted by the fact that crystallization of purified intact VSG regularly leads to crystals that contain only the NTD\cite{Freymann1984,Metcalf1988}. VSG NTDs consist of 350--400 residues and are grouped into two classes (type A and B) by the number and position of conserved cysteine residues\cite{Carrington1991,Marcello2007}. Although the sequences vary significantly between VSGs, the tertiary structure of NTDs is generally thought to be conserved, at least within each type\cite{Freymann1990,Blum1993}. The central structural elements of the NTD, derived from the structures of A-type VSGs MITat1.2 (M1.2) and ILTat1.24 (I1.24), are a coiled coil of approximately 10 nm length that is perpendicular to the cell membrane, variable surface loops at the membrane distal end, an antiparallel $\beta$-sheet and a set of short $\alpha$-helices close to its C-terminus. The CTD has been grouped into six classes (types 1--6) based on conserved sequence features\cite{Carrington1991,Berriman2005}. Structural information of the type 2 CTD of M1.2 and the type 1 CTD of I1.24 suggests that the VSG CTDs consist of one or two structured regions (S or S1 and S2) of 20--40 residues each flanked by unstructured segments (L1 and L2)\cite{Chattopadhyay2005,Jones2008}. The structured regions have a compact fold that consists of two short helices, a short antiparallel $\beta$-sheet and an optional C-terminal $\alpha$-helix.

We used small-angle X-ray scattering (SAXS) to combine newly solved high-resolution domain structures of MITat1.1 (M1.1) to provide a model for the first experimentally derived structure of a complete VSG. Additional SAXS experiments on VSG I1.24 in combination with its published high-resolution domain structures\cite{Blum1993,Jones2008} allowed a comparison between two complete VSGs, their overall architecture and the structure--function relationship of the surface coat of \textit{T.~brucei}. The results obtained by structural analyses were supported by measuring the diffusion dynamics of VSG M1.1 in artificial membranes.

\section{Results}

\paragraph{Crystal structure of the NTD of VSG M1.1.}
VSG M1.1 has an A-type NTD and a type 2 CTD, and is consequently labelled an A2 VSG (Fig.~\ref{f:NMB1:CrystalStructure}a shows the domain composition). Full-length M1.1 (sequence shown in Supplementary Fig.~1a) was purified in its soluble form (sVSG) from trypanosome cells by cleavage of the GPI anchor. In accordance with observations reported for M1.2 and I1.24, the CTD was not present in the crystals, having presumably been cleaved off by the action of endogenous proteases on the L1 region\cite{Freymann1984,Metcalf1988} (Supplementary Fig.~2).

Crystal-structure analysis (Supplementary Table 1) showed that the NTD forms a homodimer (Fig.~\ref{f:NMB1:CrystalStructure}b). Central to the structure of each monomer of the M1.1 NTD is a coiled coil formed by two anti-parallel $\alpha$-helices that span a distance of 9.4 nm (Fig.~\ref{f:NMB1:CrystalStructure}c). The coiled coil is flanked by shorter $\alpha$-helices that extend perpendicular to the central coil and represent the broadest part of the domain. This motif is 6.4 nm wide in the dimer and is positioned at a vertical distance of 2.0 nm from the C-terminal end of the domain. The structure close to the N-terminal end forms variable surface loops that span an area of 17 nm${}^2$ per dimer. Four conserved cysteine residues form disulfide bonds between residues C15 and C145 and residues C123 and C190 (Fig.~\ref{f:NMB1:CrystalStructure}b,c), which is consistent with the configuration found in the NTDs of VSGs M1.4, M1.2 and I1.24 (refs \cite{Freymann1990,Blum1993,Allen1983}).  Distance-based similarity measures were determined by superimposing M1.1 and M1.2 over 264 C$\alpha$  pairs and showed a low root mean squared deviation (r.m.s.d.) of 0.9 {\AA}, whereas superimposing M1.1 and I1.24 over 84 C$\alpha$  pairs led to an r.m.s.d. of 1.3 {\AA}, despite sequence identities of only 27.1\% and 10.6\%, respectively (all the values were calculated with UCSF Chimera MatchMaker\cite{Pettersen2004}). To highlight the similarities and differences between these three VSG NTD structures they are displayed in Supplementary Fig.~3a. An oligosaccharide of the type (Man)${}_n$(GlcNAc)${}_2$ is linked to residue N266 (Fig.~\ref{f:NMB1:CrystalStructure}b,c) with three mannose residues clearly identified from the electron--density maps. 
\begin{figure}[h!]
\includegraphics{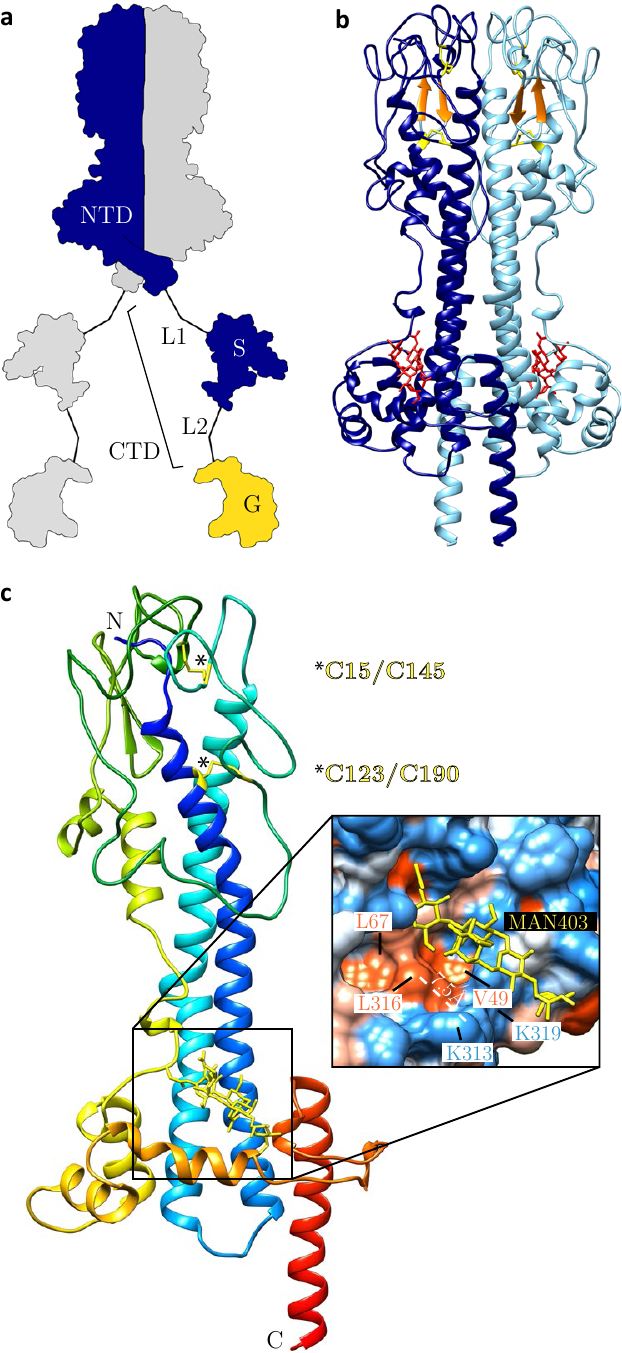}
\caption{\label{f:NMB1:CrystalStructure}\textsf{\footnotesize{\textbf{Crystal structure of the NTD of VSG M1.1.} \textbf{a}, Schematic overview of the domain architecture of the M1.1 dimer. In each monomer the NTD is connected through L1 to the structured region S of the CTD, which is in turn connected via L2 to the GPI anchor G. Domains L1 through L2 form the CTD (bracket). \textbf{b}, The NTD of M1.1 forms a rigid dimer (one monomer is coloured in dark blue and the other in light blue). The oligosaccharide at N266 is highlighted in red, the antiparallel $\beta$-sheet in orange and the disulfide-forming cysteines in yellow. \textbf{c}, One monomer of the M1.1 NTD coloured in a rainbow scheme from blue (N-terminus) to red (C-terminus). The structure includes a long coiled coil (blue, turquoise) encircled by shorter $\alpha $-helices near the C-terminal end (orange), a longer $\alpha$-helix that connects the NTD to the CTD (red) and variable surface loops near the N-terminal end (green). The two cysteine pairs that form disulfide bonds are marked with asterisks. The assignable part of the glycan, (Man)${}_3$(GlcNAc)${}_2$, which is covalently attached to residue N266, is shown in yellow. Inset: a hydrophobic pocket is formed around L316 and is shielded by the oligosaccharide. The distance to the nearest neighbouring surface residue K313 is 7.5 {\AA}.}}}
\end{figure}
An additional signal in the electron--density map for the inherently flexible and heterogeneous oligosaccharide could not be resolved, but suggests the presence of a larger glycan, most probably a (Man)${}_{4-9}$(GlcNAc)${}_2$ composition, which is common among VSGs, for example, (Man)${}_4$(GlcNAc)${}_2$ in M1.2 (refs \cite{Zamze1990,Bangs1988,Strang1993}). 
\begin{figure*}[ht]
\includegraphics{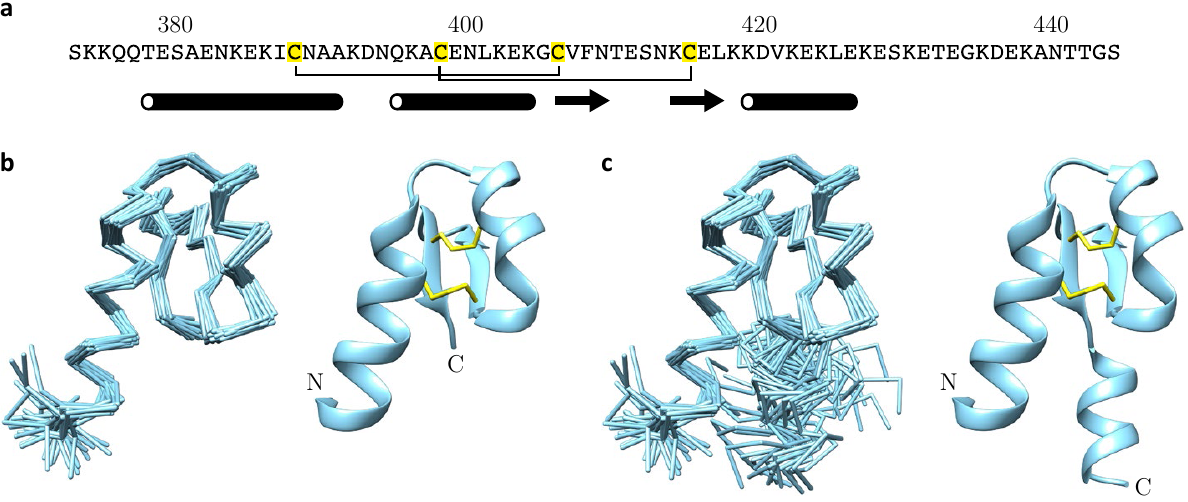}
\caption{\label{f:NMB2:SolutionStructure}\textsf{\footnotesize{\textbf{Solution structure of the CTD of VSG M1.1.} \textbf{a}, Amino acid sequence of the M1.1 CTD, residues S372 to S443, with the four conserved cysteines highlighted in yellow and disulfide pairing indicated. The location of the experimentally determined secondary structure elements is shown underneath the sequence (barrels for $\alpha$-helices and arrows for $\beta$-strands). \textbf{b}, Overlay of residues Q376 to K417 of the 35 lowest-energy structures (left) and the structure closest to the mean displayed as a ribbon diagram with disulfide bonds shown in yellow (right). \textbf{c}, Overlay of residues Q376 to K426 of the 35 lowest-energy structures (left) and the structure closest to the mean displayed as a ribbon diagram with disulfide bonds shown in yellow (right). }}}
\end{figure*}
The glycan is positioned above a hydrophobic pocket formed by V49, L67, L316 and K319 that is recessed by at least 7.5 {\AA} from the most proximate residue on the protein surface, K313 (Fig.~\ref{f:NMB1:CrystalStructure}c, inset).

\paragraph{Solution structure of the CTD of VSG M1.1.}
The CTD used to elucidate the structure of VSG M1.1 was expressed recombinantly. The polypeptide contained residues S372 to S443 of the mature protein with an added glycine at the N-terminus. This residue was a remnant from the removal of the affinity tag used for purification (Supplementary Fig.~1a,b).

The 35 lowest-energy structures generated (Supplementary Table 2) showed that the CTD structure is composed of two $\alpha$-helices separated by a turn and a short antiparallel $\beta$-sheet, with two disulfide bonds between C387 and C405 and between C397 and C414 fixing these four secondary structure elements into a compact arrangement (Fig.~\ref{f:NMB2:SolutionStructure}a,b). In addition to this core structure, there is a short $\alpha$-helix following the antiparallel $\beta$-sheet (Fig.~\ref{f:NMB2:SolutionStructure}c). The C-terminal $\alpha$-helix is defined by its short-range nuclear Overhauser effect (NOE) pattern and the positive (albeit small) heteronuclear NOE values, although no long-range NOEs were observed between the core structured region and this C-terminal $\alpha$-helix. Consequently, every calculated structure for M1.1 contains the C-terminal $\alpha$-helix, but the overlay of 35 structures clearly shows that this helix is not anchored to the rest of the structure. The helix is therefore locally ordered, but its position with respect to the rest of the domain is not well defined (Fig.~\ref{f:NMB2:SolutionStructure}c). A comparison of the structure of the CTD of M1.1 with the previously reported CTD structures of M1.2 and I1.24 (S1 domain only) is shown in Supplementary Fig.~3b.

\begin{figure*}[ht]
\includegraphics[width=0.794\textwidth]{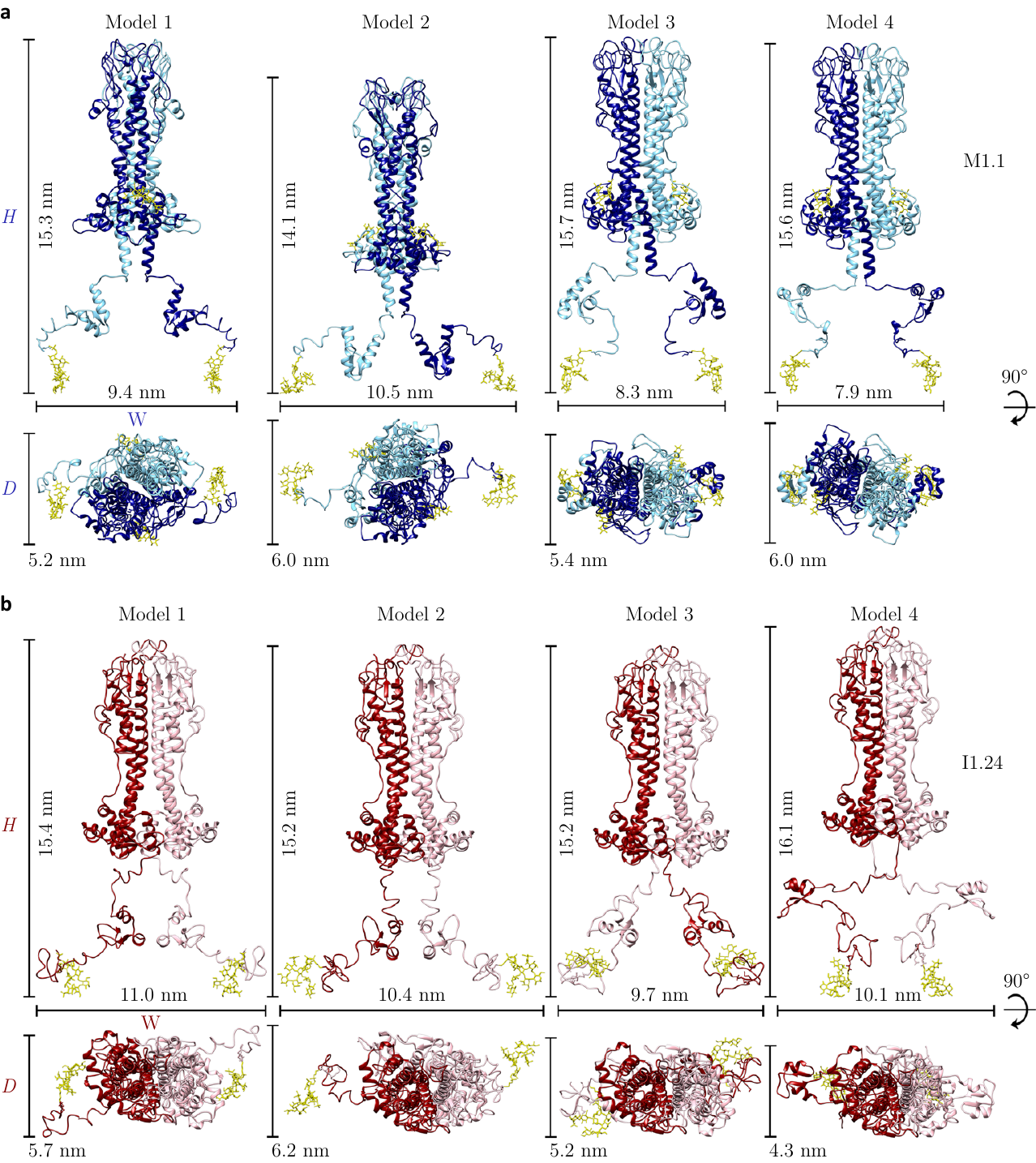}
\caption{\label{f:NMB3:Models}\textsf{\footnotesize{\textbf{Rigid-body models of two complete VSGs.} \textbf{a}, Side and top-down view of four models (1--4) of VSG M1.1. The protein dimensions parallel (width, $W$, and depth, $D$) and perpendicular (height, $H$) to the plane of the plasma membrane are displayed. The two monomers that comprise each dimer are shown in light and dark blue. The N-glycan and GPI anchor are displayed in yellow. \textbf{b}, Side and top-down view of four models (1--4) of VSG I1.24. The two monomers of each dimer are shown in light and dark red. Bar lengths are displayed according to the model dimension.}}}
\end{figure*}

\paragraph{The structure of VSG M1.1 adopts two main conformations.}
SAXS of M1.1 sVSG was used to generate a low-resolution molecular envelope for the complete molecule (Supplementary Fig.~ 4c). This envelope was then used to guide the placement of the high-resolution structures of the NTD and CTD into a complete VSG structure. The CTD of each monomer features a single structured domain S that is connected to the NTD and the GPI anchor by the flanking unstructured linkers L1 and L2, respectively (Fig.~\ref{f:NMB1:CrystalStructure}a). This is the simplest CTD composition in \textit{T.~brucei} VSGs and represents a good starting point for the analysis of a complete VSG. Rigid body modelling resulted in a set of four models with a good fit to the experimental data, but with different spatial arrangements of the CTD relative to the NTD (models in Fig.~\ref{f:NMB3:Models}a and Supplementary Figs 4 and 5). The distance spanned by the two CTDs in a VSG dimer varies between 7.9 nm and 10.5 nm in the four models (Fig.~\ref{f:NMB3:Models}a) with model 2 possessing the greatest width. Consequently, its long axis of 14.1 nm is smaller than the average of 15.5 nm for the other models (Supplementary Table 3).

\begin{figure*}[ht]
\includegraphics{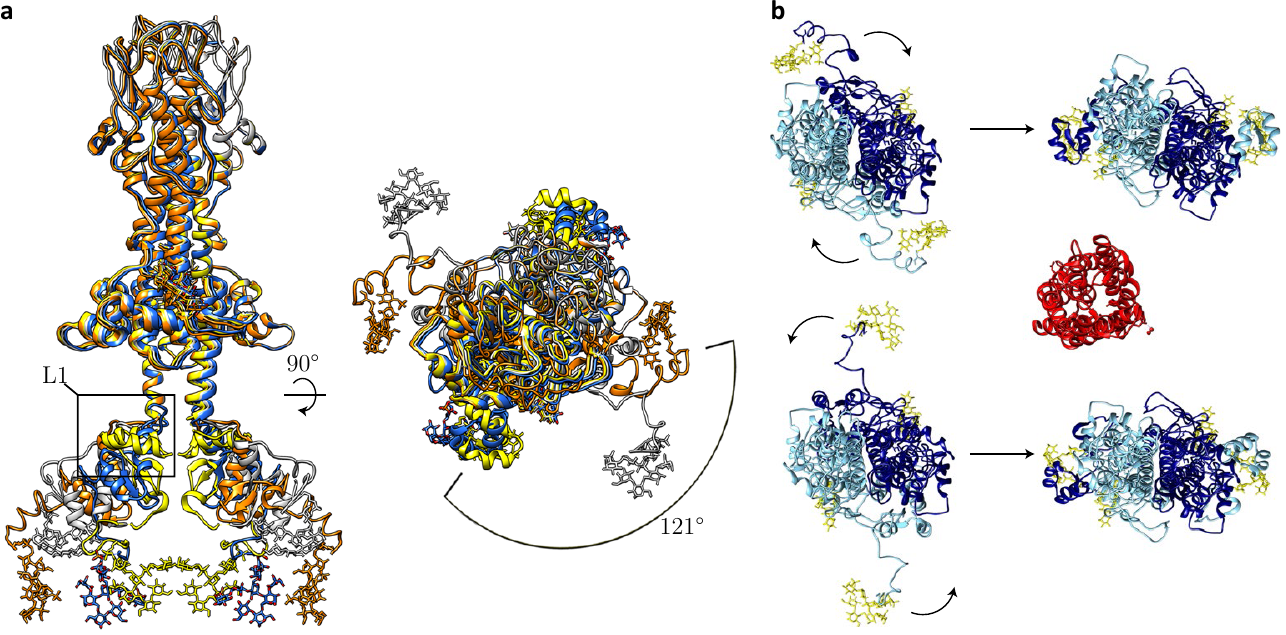}
\caption{\label{f:NMB4:Flexibility}\textsf{\footnotesize{\textbf{Flexibility of linker L1 in VSG M1.1.} \textbf{a}, The superimposed NTDs of models 1--4 of M1.1 demonstrate the flexibility of the CTD, which varies in position along the interdomain axis by up to 121${}^{\circ }$ facilitated through linker L1. Supplementary Video 1 gives a more detailed view of the CTD localization. \textbf{b}, Top-down illustration of VSGs (blue) with a surface transmembrane protein (aquaporin 2 (PDB ID: \href{https://www.rcsb.org/structure/1FX8}{1FX8}, red)) shown to scale. The flexibility in linker L1 may allow the CTD to respond to obstacles, for example, transmembrane proteins, while the NTD position remains unaffected.}}}
\end{figure*}

Superimposing the NTD of the four initial models displays a shift in the placement of the CTD. The position of the CTD varies by up to 121${}^\circ$ along the interdomain axis for the representative models (Fig.~\ref{f:NMB4:Flexibility}a and Supplementary Video 1), which suggests that linker L1 confers interdomain freedom.

Not taking into account the hydration shell present on the protein surface and presuming a rigid conformation, the area a single VSG dimer theoretically occupies varies between at least 25 nm${}^2$ and 31 nm${}^2$, allowing a total of $3.2-4.0 \times 10^4$ VSGs to fit into an area of 1 \musq (Supplementary Fig.~6 and Supplementary Table 3). Assuming 144 \musq as the surface area of a single cell\cite{Gruenfelder2002}, the calculated values would amount to $4.6-5.8 \times 10^6$ VSGs per cell, which is in good agreement with the published $5.7 \times 10^6$ VSG dimers per cell\cite{Jackson1985}. However, these values assume a static protein equally distributed on the cell surface and underestimate the actual area occupied by VSGs. To account for the rotational freedom between the CTD and the NTD as well as the rotational diffusion of the protein on the cell surface, the circular area was calculated using the distance spanned by the two CTDs in the VSG dimer with an added theoretical hydration layer of 1 nm as the diameter. This results in a total area per VSG dimer of 77--123 nm${}^2$, allowing $0.8-1.3 \times 10^4$ VSGs to diffuse freely in an area of 1 \musq, which corresponds to $1.2-1.9 \times  10^6$ VSG dimers per parasite.

All the preceding M1.1 models were computed using two-fold symmetry and with contact conditions for the N-terminal dimer to approximate the restrictions conferred by the cell membrane. To allow for the greater degree of freedom available for VSGs in solution, additional models were calculated without specifying the symmetry of the protein or contact conditions for the NTD. The resulting structures displayed correctly dimerized NTDs and an asymmetric conformation of the CTD (Supplementary Fig.~ 4b). Although the CTD of chain A extends in parallel to the symmetry axis, that of chain B is tilted back towards the NTD.

\paragraph{The structure of VSG I1.24 also adopts two main conformations.}
I1.24, like M1.2, possesses an A-type NTD, but its CTD is of type 1. This is a more-complex type of CTD composed of two structured regions S1 and S2. A flexible linker (L1) connects the NTD to S1 with another such linker (L2) that connects S1 to S2, which is directly connected to the GPI anchor. To compare the influence of different types of CTDs on the overall VSG structure, SAXS measurements and rigid-body modelling of I1.24 were conducted in parallel to M1.1. 
The CTD of I1.24 consists of 109 residues, compared with the 75 residues of the M1.1 type 2 CTD. Previously published high-resolution data of the NTDs and CTDs of I1.24 (refs \cite{Blum1993,Jones2008}) (ILTat1.24; Protein Database (PDB) ID: \href{https://www.rcsb.org/structure/2VSG}{2VSG} and \href{https://www.rcsb.org/structure/2JWG}{2JWG}, \href{https://www.rcsb.org/structure/2JWH}{2JWH}) were used to deduce the complete VSG structure, which resulted in four models (1--4) (Fig.~\ref{f:NMB3:Models}b and Supplementary Table 3). With a width of 9.7--11.0 nm, I1.24 spans a slightly greater distance than M1.1, whereas the average height is slightly shorter at 15.5 nm. However, both proteins show similar compact and extended conformations. The area spanned by the I1.24 dimers ranges from 25 nm${}^2$ to 35 nm${}^2$ for non-hydrated static models, which could form a coat composed of $2.9-4.0 \times  10^4$ VSG dimers per square micrometre. The circular area occupied by I1.24 is calculated from the maximum width of the hydrated protein and spans $107-133$ nm${}^2$. This corresponds to $0.8-0.9 \times  10^4$ VSGs $\mu$m${}^{-2}$. Compared with M1.1, $10-31\%$ fewer I1.24 VSG molecules are required to pack the cell surface densely.

\paragraph{Two distinguishable freely diffusing populations of VSG M1.1 are observed in supported lipid bilayers.}
The above SAXS experiments were performed on VSGs in solution, in which the proteins experience fewer constraints than when anchored to the plasma membrane. As changes in the conformation of the VSG protein might well manifest themselves as differences in diffusion behaviour, we performed single-molecule diffusion experiments on membrane-bound VSG M1.1. A well-defined solid-supported lipid bilayer system that guaranteed a uniform membrane composition and the presence of only one protein species, the membrane form of M1.1 (Supplementary Fig.~7), was chosen (Methods).

In addition, the low concentration of proteins $(0.0897 \pm 0.0003 \textrm{ } \mu\mathrm{m}^{-2})$ used in this experiment ensured diffusion measurements of single M1.1 mfVSG molecules. The quality of the lipid bilayer was validated by analysing the diffusion of individual lipids. After excluding data that showed non-Brownian statistics from the 1,890 lipid traces acquired (minimum and average trace length of nine and 13, respectively), a single lipid population that exhibited Brownian motion was found. A diffusion constant of $ D_\mathrm{lipids} =  2.688 \pm 0.005 \textrm{ } \mu \mathrm{m}^2 \textrm{ s}^{-1}$ was obtained from time averaging (Methods). The uncertainty here and henceforth is the standard error, unless stated otherwise.
\begin{figure*}
\includegraphics[width=\textwidth]{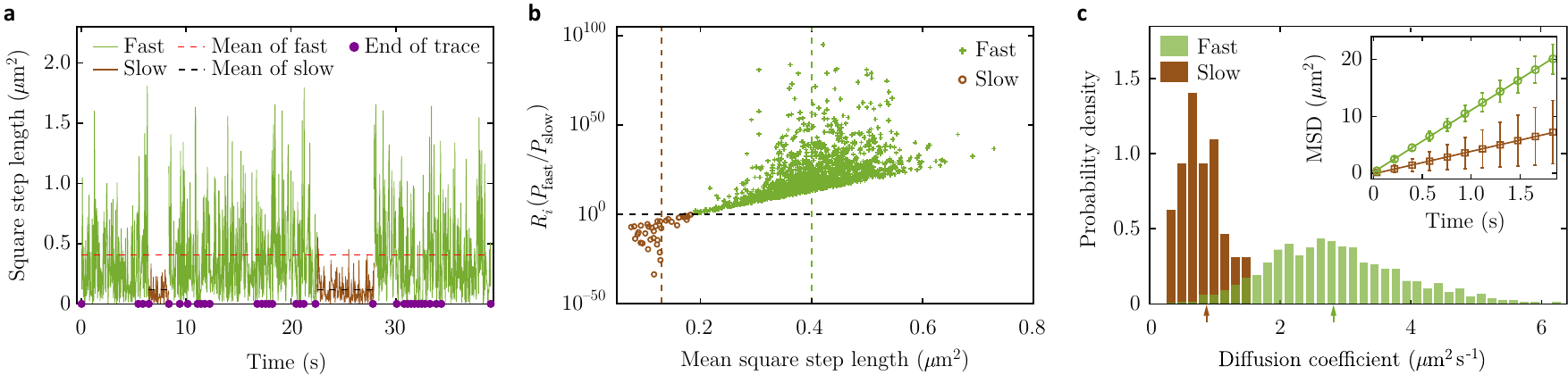}
\caption{\label{f:NMB5:DiffusionBehaviour}\textsf{\footnotesize{\textbf{Diffusion behaviour of mfVSG M1.1 incorporated into a solid supported lipid bilayer.} \textbf{a--c}, Analysis of single-molecule diffusion data reveals a small population (3.54\%) of slowly diffusive proteins (brown) and a large population (96.46\%) that is significantly faster (green). \textbf{a}, The square of measured step lengths (775 in total) as a function of time in a representative set of 32 traces associated with fast and slow processes, after filtering for Brownian diffusion. Individual traces are separated by purple dots on the abscissa. The mean square step length is indicated with a dashed-red (fast) and dashed-black (slow) line. \textbf{b}, Scatter plot that shows the separation of traces into two populations. Each of the 1,602 traces is represented by a point given by the mean square step length in the trace (abscissa), and the ratio of likelihoods for a trace to belong to the fast and the slow diffusing population (ordinate). The mean square step lengths for the two populations are indicated with vertical dashed lines. The horizontal dashed line indicates equal likelihood. \textbf{c}, A histogram that displays the distribution of diffusion coefficients for the population of fast and slow proteins shown in \textbf{b}. The inset shows the mean square displacement as a function of time, calculated from the cumulative trace of the two populations. Every fifth lag-time point is displayed. Error bars are standard errors. The mean diffusion coefficients are indicated by arrows. The data displayed in \textbf{b} and \textbf{c} are representative of five technical replica measurements.}}}
\end{figure*}

Next, the diffusive behaviour of individual VSG molecules was measured. A total of 2,969 trajectories were collected, with a minimum and average trace length of 20 and 49, respectively.

Non-Gaussian stretches of protein trajectories were filtered and excluded from analysis. The immobile fraction of the VSG was in the same order of magnitude as for the lipids (Supplementary Fig.~8).

All the remaining traces performed Brownian motion, with separate mean square step lengths and variance (brown and green in Fig.~\ref{f:NMB5:DiffusionBehaviour}a, and Supplementary Fig.~9). Two families clearly separated with the ratio of likelihoods to belong to either family used as an additional sorting criterion. Following this grouping into families of slow (3.54\%) and fast (96.46\%) traces (Fig.~\ref{f:NMB5:DiffusionBehaviour}b), probability distributions of the diffusion coefficients were generated by independently fitting the time evolution of mean square displacement for each trace (Fig.~\ref{f:NMB5:DiffusionBehaviour}c). The average over the ensemble of these diffusion coefficients provided the mean coefficients for fast and slow diffusion, which were $2.87 \pm 0.03(1.08) \textrm{ } \mu \mathrm{m}^2 \textrm{ s}^{-1}$ and $0.81 \pm  0.05(0.3) \textrm{ } \mu \mathrm{m}^2 \textrm{ s}^{-1} $, respectively. The numbers in brackets comprise the width or standard deviation of the distribution. As expected for Brownian diffusion, these results corresponded well to the diffusion coefficients obtained from time averaging, in which a single mean square displacement curve was constructed from concatenated trajectories. Here, the respective diffusion coefficients were $2.716 \pm  0.005 \textrm{ } \mu \mathrm{m}^2 \textrm{ s}^{-1}$ and $0.848 \pm  0.002 \textrm{ } \mu \mathrm{m}^2 \textrm{ s}^{-1}$ (Fig.~\ref{f:NMB5:DiffusionBehaviour}c). Notably, the diffusion of the fast family was very similar to the diffusion of the lipids.

The overall statistical analysis performed here shows that two distinct populations of proteins exist in the membrane. Both families exhibited Brownian diffusion with coefficients that were in the same order of magnitude. These diffusion coefficients are, however, significantly distinct, which is reflective of our structural findings that two major VSG conformations exist.

\section{Discussion}

\begin{figure*}
\includegraphics{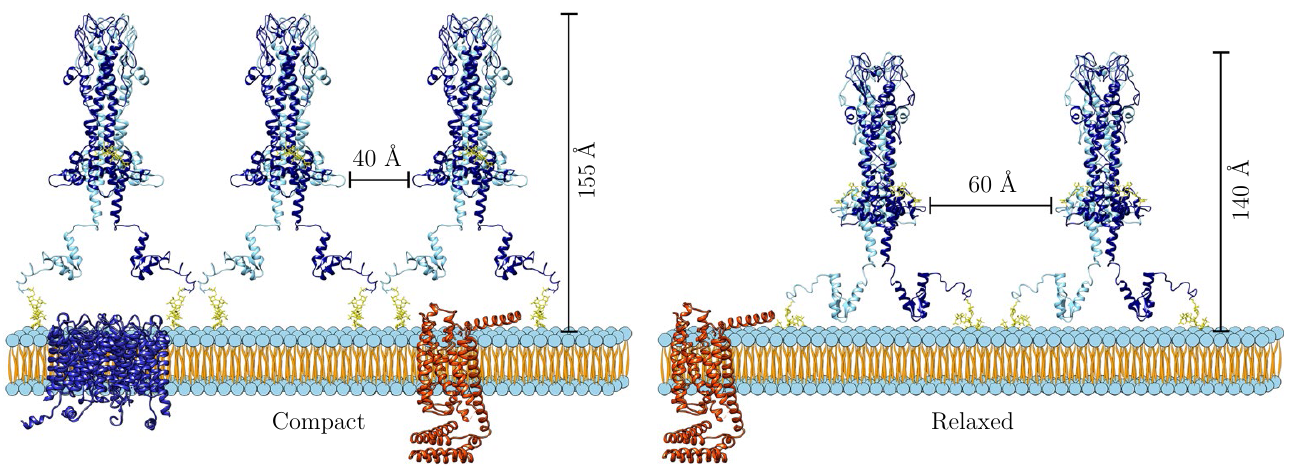}
\caption{\label{f:NMB6:VSGPacking}\textsf{\footnotesize{\textbf{VSG packing on the cell surface.} Side-view illustration of membranes occupied by VSGs at two different densities. The conformation of tightly packed VSGs (left) can elevate the VSG above the transmembrane proteins (dark blue, red), whereas the relaxed conformation (right) could allow the maintenance of a protective coat on the cell surface even at a reduced protein density.}}}
\end{figure*}

In the present study, we solved the tertiary structure of the NTDs and CTDs of a third VSG, M1.1 (A2 VSG), which is largely in agreement with the previously reported structures of M1.2 (A2 VSG) and I1.24 (A1 VSG). We also report the complete structures of two VSGs that were obtained by combining the high-resolution X-ray crystal structures of the NTD and the high-resolution NMR structures of the CTD with low-resolution SAXS experimental data. SAXS measurements are performed with proteins in solution, in which the molecules can freely adopt a variety of conformations. Under these conditions, they do not experience the geometric constraints imposed, for example, by GPI anchoring into the plane of a two-dimensional (2D) membrane space. The constraints enforced on the dimeric protein by dimerization of the NTD, on the one hand, and insertion of the lipid moiety of the GPI anchor into the membrane, on the other hand, suggest that the two CTDs in the dimer must adopt similar conformations. We therefore computed our models in two ways: (1) without restraints, which mimics the conditions found in solution, and (2) by defining a two-fold symmetry of the VSG as an approximation of the conditions found when naturally anchored to the lipid membrane. Structures obtained using this filtering step are a subset of the possible conformations in solution. All the models generated using these two approaches reassuringly showed a dimerized NTD. The CTDs of a dimer in the models generated without filtering for symmetric VSGs could, however, position asymmetrically (Supplementary Fig.~4b). This result emphasizes the high degree of flexibility of the VSG, which is mediated by linker L1. Filtering for conformations that conform to the restraints imposed on the molecule when anchored to the membrane resulted in models in which the CTDs and GPI anchors fold towards the presumed membrane location. This means that the environment of the VSG when bound to the membrane significantly contributes to the overall VSG structure. The flexibility conferred by L1 is evident here, too, from the different conformations a symmetrical VSG can adopt. These can be grouped into two main folds. One fold is extended and slim (compact conformation), and the other extends less from the membrane and has a wider footprint (relaxed conformation) (Fig.~\ref{f:NMB6:VSGPacking}). The individual SAXS models generated for one VSG differ by up to 30\% in the area they occupy. This variance does not contradict a conserved structure--function relationship because it is true for the model pool of both the analysed VSGs. Instead, we propose that these individual models display conformations that can alter dynamically in response to changes in the overall protein density on the membrane. This hypothesis is supported by the presence of two freely diffusing populations of VSG in an artificial lipid bilayer system. This may allow the protein to adapt its structure to remain functional under different environmental conditions encountered both on the cell surface and when present in intracellular compartments during recycling\cite{Gruenfelder2003,Overath2004}.

Antigenic variation replaces an old VSG with a new one. The present data suggest that individual VSGs may occupy a range of different surface areas on the plasma membrane. However, the shielding function of the VSG must not be compromised at any time. Here the flexible conformations of the VSGs, which allow an adaptation to a range of different protein densities, may extend the protein's capability to function in different environmental conditions. We suggest that continuous transient variation between the two main VSG conformers could compensate for such natural fluctuations and imprecisions, and provide robustness through structure to the VSG function.

We have previously examined the maximum protein density of the VSG coat\cite{Hartel2015,Hartel2016}. The simultaneous activation of two independent VSGs in transgenic trypanosomes leads to an overexpression of up to 150\% of the total VSG\cite{Batram2014}, which may force all VSGs to adapt a compact conformation. The removal of the C-terminal oligosaccharide drastically reduced the la\-te\-ral mobility of VSG M1.6 in artificial bilayers\cite{Hartel2016}. This work on VSG dynamics also showed that the natural VSG density in a defined membrane can be increased by approximately 40--50\% before the molecular crowding threshold (MCT) is reached and the diffusion coefficient and mobile fraction of the coat break down. We suggest that additional VSG on the membrane results in a conformational change towards a compact conformation and that only once all the VSG molecules have adopted a compact state does the surface coat reach its spatial limitations and the mobility collapses on a further increase of protein density. This MCT-dependent collapse is probably connected to protein deformations beyond the normal range of conformations. In the SAXS experiments reported here, the VSG was not in such a highly crowded environment.

Although VSGs dominate the trypanosome cell surface, they are not the only proteins to populate the plasma membrane. Like all eukaryotes, trypanosomes rely on receptors and channels for nutrient uptake, sensing and communication. These invariant transmembrane proteins may affect the free diffusion of VSGs. High mobility, however, is essential for VSG coat function. We suggest that the compact form of the VSG lifts the protein well above the membrane and allows floating over, for example, channels and receptors up to the size of a G-protein-coupled receptor (Fig.~\ref{f:NMB6:VSGPacking}). We envisage the CTD to function like the suspension system of a road vehicle, cushioning the influence of obstacles, such as invariant surface proteins, to allow the NTD to remain unperturbed (Fig.~\ref{f:NMB4:Flexibility}a,b). It is also possible that the VSG deforms on binding of the host antibodies. IgG and IgM would tether two VSGs and the NTD---through L1---may bend due to the flow forces during antibody clearance, while the rest of the CTD ensures the necessary mobility.

Thus, the unique feature of functional importance that sets VSGs apart from structurally similar proteins\cite{Salmon1997,Lane-Serff2014} is the presence and flexibility of its CTD.

This is reflected by the fact that the total width of the VSG is always determined by the CTD (Fig.~\ref{f:NMB3:Models}), which spreads out furthest towards the surrounding protein coat in our models. The average width of the M1.1 and I1.24 models is 9.0 nm${}^2$ and 10.3 nm${}^2$, respectively, and the average area they occupy for a non-hydrated and static model is 28 nm${}^2$ and 30 nm${}^2$, respectively. So, despite a sizable difference, of more than 30\%, in the CTD of these two VSGs, the average area that each VSG spans varies by only 7\% (Supplementary Table 3). However, when considering the rotational freedom between the CTD and NTD and the high mobile fraction of VSGs on the cell surface\cite{Hartel2016}, the static model obviously underestimates the actual area each VSG occupies. The consideration of a highly dynamic VSG molecule results in an extra 20\% average area spanned by I1.24 owing to the increased size of its CTD. Individual SAXS models display a variance of up to 30\% in the area they occupy. This means that different VSGs can occupy different areas and in turn, the number of VSG molecules required to cover the cell surface is not constant. The only experimental analysis with respect to the number of VSG molecules on the trypanosome cell surface was published in 1985 by Jackson et al.\cite{Jackson1985}. These authors determined the number of VSG molecules to be about $5\times 10^6$ dimers per cell. A direct comparison of the VSG molecules per trypanosome from different experiments is difficult as it has recently become clear that trypanosome morphology is fairly heterogeneous\cite{Bargul2016}, and we therefore present VSG numbers normalized to a defined surface area of 1 $\mu$m${}^2$. Nevertheless, the commonly cited $5\times 10^6$ dimers per cell is in good agreement with the computed number of M1.1 VSGs based on our SAXS data.

In summary, VSGs adopt two main conformations that appear to allow the shielding function to be maintained at varying protein densities, and also enable a high mobility in the presence of obstacles and confinement, and protecting receptors from immune detection. This molecular shield is not a static barrier, but highly mobile and structurally amazingly flexible. Beside trypanosomes, many other eukaryotes possess a GPI-anchored surface coat, not all of which are parasites. It would be interesting to see if those cells have evolved similar structure--function relationships of their major surface proteins.

\section{Methods}

\paragraph{Generation of a transgenic I1.24-expressing cell line.}
The open reading frame of VSG I1.24 (IL.24 (UniProt accession number P26329)) and its 3'-UTR up to and including the conserved 14-mer sequence was cloned into the EcoRI site of the M1.2ES targeting vector pKD (ref.\ \cite{Munoz-Jordan1996}). Following linearization with BglII, the construct was integrated into the genome of a M1.2 expressing \textit{T.~brucei} Lister 427 strain to yield a cell line that expressed both VSG M1.2 and I1.24. Knockout of the M1.2 gene by replacement with a puromycin cassette led to the generation of transgenic cells that expressed I1.24 from the active M1.2ES.

\paragraph{Purification of sVSG from \textit{T.~brucei}.}
The VSG is the most abundantly expressed protein in \textit{T.~brucei} and constitutes ~10\% of the total protein\cite{Boehme2002}. It can be readily purified directly from trypanosomes after release by endogenous GPI-specific phospholipase C (GPI-PLC)\cite{Cross1984,Ferguson1985}. \textit{T.~brucei} wild-type strain Lister 427, which expresses VSG M1.1 (MITat1.1; UniProt accession number \href{http://www.uniprot.org/uniprot/P26331}{P26331} with amino acid changes K31E and S65L), or a transgenic Lister 427 cell line that expresses VSG I1.24 from the M1.2 ES was used to purify the respective VSGs. Trypanosomes were cultivated in HMI-9 medium\cite{Hirumi1989} with 10\% fetal bovine serum (Sigma-Aldrich) at 37\degc{} with 5\% CO${}_2$ using conical flasks agitated on an orbital shaker at 40--80 r.p.m. to allow growth past the conventionally obtained maximum of $4\times 10^6$ cells ml${}^{-1}$. Approximately $5\times 10^9$ cells were harvested by centrifugation at 5,000$g$ for 30 min at 4\degc{} and washed in PBS. GPI-anchored proteins were released by endogenous GPI-PLC after the induction of osmotic shock in 10 mM sodium phosphate buffer, pH 8.0 (ref.\ \cite{Cross1984}). The hypotonic buffer was supplemented with 0.1 mM tosyl-\textsc{l}-lysyl-chloromethane hydrochloride, 7.5 $\mu$M leupeptin and 0.5 $\mu$M aprotinin to protect the sVSG from proteolytic degradation by cellular proteases. The buffer was changed to 20 mM Tris-HCl, pH 8.0, using HiTrap desalting columns (GE Healthcare) on an {\"A}ktaprime plus (GE Healthcare). Purification of full-length VSG from the lysate was conducted by anion-exchange chromatography using HITrapQ HP columns (GE Healthcare) in 20 mM Tris-HCl, pH 8.0 with a 0--1 M NaCl gradient. sVSG eluted from the anion-exchange column at approximately 400 mM NaCl and VSG positive fractions were pooled. Subsequent purification and removal of the aggregated sVSG was performed by size exclusion chromatography on a HiLoad 16/600 Superdex 200 pg column (GE Healthcare) in 20 mM Tris-HCl, pH 8.0, 150 mM NaCl. Fractions that contained purified protein were pooled again, concentrated and the buffer was changed to H${}_2$O using Amicon Ultra-15 and Ultra-0.5 centricons (Millipore) with a cutoff of 10--30 kDa at 5,000$g$ and 4\degc{} for 10 min. Purified VSG was flash-frozen in liquid nitrogen and stored at $-20$\degc{}. This method was used to purify VSGs for both macromolecular crystallography and SAXS experiments. 

\paragraph{Purification of the recombinant M1.1 CTD. }
For the structural analysis of the CTD of VSG M1.1 (MITat1.1; UniProt accession number \href{http://www.uniprot.org/uniprot/P26331}{P26331}), a construct was designed to express residues S372 to S443 of the mature protein. A tobacco etch virus (TEV) protease recognition site (ENLYFQG) was added to the open  reading frame and cloned into the XhoI and BamHI sites of pET15b (Novagen) to produce a polypeptide with a TEV protease cleavable N-terminal $6\times $His-tag (MGSSHHHHHHSSGLVRPGSHMLEENLYFQG). Both an unlabelled and a ${}^{15}$N-labelled polypeptide were expressed in \textit{Escherichia coli} BL21trxB(DE3), grown in $2\times $TY medium or minimal medium based on MOPS buffer that contained 5\% ${}^{15}$N-labelled Celtone and [${}^{15}$N] ammonium chloride (Spectra Stable Isotopes), respectively. After cell lysis, the polypeptide was isolated from the crude protein extract by affinity chromatography on a Ni${}^{2+}$ column. The $6\times $His-tag was subsequently removed by digestion with TEV protease to leave an additional Gly residue at the N-terminus of the M1.1 CTD. Final purification was performed by reversed-phase HPLC over a $25\times 1$ cm Hichrom-5 C8 column in an acetonitrile gradient in the presence of 0.1\% trifluoroacetic acid.

To confirm the protein's identity and disulfide status, the relative molecular mass of the unlabelled M1.1 CTD was determined via electrospray ion trap mass spectrometry. The mass determined for the M1.1 CTD was 8,135.77 Da, which is in close agreement with the expected mass of 8,135.0 Da for the fully disulfide-bonded form. Under reducing conditions, the mass increased to 8,138.87 Da with the expected mass of the fully reduced polypeptide being 8,139.0 Da.

\paragraph{Crystallization, structure determination and refinement.}
Crystal screens were performed using the Honeybee 963 crystallization robot (Zinsser Analytic). Crystal growth could be observed between 14 and 40 days of incubation using the sitting drop vapour-diffusion method at 20\degc{} with a protein concentration of 20 mg ml${}^{-1}$ in H${}_2$O mixed in a 1:1 ratio with 0.1 M MES buffer, pH 5.5--6.5, 13--18\% poly(ethylene glycol) (PEG) 20,000. Crystals were cryoprotected by brief incubation in 0.1 M MES, pH 5.5--6.5, 15--20\% PEG 20,000, 25\% glycerol, and subsequently flash-frozen and stored in liquid nitrogen in CryoLoops (Hampton Research).

Diffraction measurements were conducted at beamline MX 14.1 at the Berliner Elektronenspeicherring-Gesellschaft f{\"u}r Synchrotronstrahlung (BESSY II), using a PILATUS 6M detector (Dectris) to collect the diffraction data from a single crystal mounted in a cold nitrogen-gas stream\cite{Gerlach2016}. 

M1.1 crystals were obtained with unit-cell dimensions of $a=78.5$ {\AA}, b =  95.0 {\AA} and $c =  103.8$ {\AA} in space group $P2_1 2_1 2_1$. The crystals diffracted to a resolution of 1.65 {\AA} (Supplementary Table 1) and contained amino acids A1 to L368 of the mature protein, which corresponds to the residues that form the NTD. In accordance with this result, SDS--PAGE analysis revealed a reduced molecular mass that corresponds to 40 kDa for the monomers of the crystallized protein compared with those of the freshly purified sVSG (55 kDa) (Supplementary Fig.~2).

The CCP4 program suite was used for structure solution and refinement unless otherwise indicated\cite{Winn2011}. Diffraction data were integrated using iMosflm\cite{Battye2011} and scaled with Scala\cite{Evans2006} or Aimless\cite{Evans2011}. Phasing was carried out by molecular replacement using the program Phaser\cite{McCoy2007} with the NTD of M1.2 (MITat1.2; PDB ID: \href{https://www.rcsb.org/structure/1VSG}{1VSG}), which shares a 27.1\% sequence identity with M1.1, as a search model. The initial automatic model building was performed using ARP/wARP\cite{Langer2008} and Buccaneer\cite{Cowtan2006}, followed by manual model building with COOT\cite{Emsley2010} and structure refinement with REFMAC\cite{Murshudov1997}.

Structural identities between VSGs were calculated with UCSF Chimera\cite{Pettersen2004}, using the Needleman--Wunsch alignment algorithm\cite{Needleman1970} and a BLOSUM-62 matrix\cite{Henikoff1992}.

\paragraph{NMR spectroscopy and solution structure determination. }
NMR samples of both the unlabelled and ${}^{15}$N-labelled polypeptide were prepared with the protein at a concentration of 0.5 mM in 50 mM sodium phosphate, pH 6.0, 150 mM sodium chloride and 10\% D${}_2 $O. Spectra were acquired at 25\degc{} on Bruker DRX500, DRX600 and DRX800 spectrometers. Experiments performed were double-quantum filtered correlation spectroscopy, total correlation spectroscopy (TOCSY) (72 ms mixing time), nuclear Overhauser effect spectroscopy (NOESY) (200 ms mixing time), ${}^{15}$N heteronuclear single quantum correlation, ${}^{15}$N-separated TOCSY (72 ms decoupling in the presence of scalar interactions mixing), ${}^{15}$N-separated NOESY (150 ms mixing time) and HNHA.

Spectra were processed with AZARA (W.~Boucher) and analysed with ANSIG\cite{Kraulis1994}. NOE, dihedral angle (from HNHA) and hydrogen-bond restraints were generated to yield a total of 1,008 unambiguous and 257 ambiguous non-degenerate distance restraints, and structure calculations were performed with ARIA1.2 (ref.\ \cite{Linge2003}) interfaced to CNS1.0 (ref.\ \cite{Linge2003}) using 90 ps of high-temperature dynamics and 30 ps of cooling from 10,000 to 1,000 K in torsion angle space. This was followed by 24 ps of cooling from 1,000 to 50 K using dynamics in Cartesian space. In the last iteration, 100 structures were calculated and the 35 lowest-energy structures were used for further analysis (Supplementary Table 2).

\paragraph{SAXS and rigid-body modelling of a complete VSG. }
SAXS data for VSGs were collected at beamline BM29 (refs \cite{Brennich2016,Pernot2013}) at the European Synchrotron Radiation Facility (ESRF). Protein serial dilutions were measured utilizing a concentration range of 0.63--3.14 mg ml${}^{-1}$ for M1.1 and 0.68--3.39 mg ml${}^{-1}$ for I1.24, and all the samples were mounted by a robotic sample changer\cite{Round2008}. The synchrotron beam was collimated to $0.7 \times 0.7$ mm at the detector and scattering data with a $q$ range of 0.025--5 nm${}^{-1}$ were recorded with a PILATUS 1M detector (Dectris) at 293 K at a sample-to-recorder distance of 2.87 m with an exposure of $10\times 1$ s (Supplementary Fig.~5 and Supplementary Table 4 give the SAXS data). The data of each protein series were scaled and extrapolated to 0 concentration after subtraction of the buffer using PRIMUS\cite{Konarev2003}. The radius of gyration $R_\mathrm{g}$ and the zero-angle scattering intensity $I(0)$ were calculated using AUTORG\cite{Petoukhov2007} and data points preceding the Guinier approximation range were cut from the extrapolated dataset. The pair-distance distribution function $P(r)$, the maximum particle dimension $D_\mathrm{max}$ and $R_\mathrm{g}$ were calculated using GNOM\cite{Svergun1992} and were consistent with the estimates from the Guinier approximation. The molecular mass was derived using the Porod invariant on the excluded volume of the hydrated particle\cite{Petoukhov2012}. Using a scattering vector of $q_\mathrm{max}=3$ nm${}^{-1}$ and the previously determined GNOM files, low-resolution molecular envelopes were constructed by averaging ten individual DAMMIN\cite{Svergun1999} models with imposed P2 symmetry by DAMAVER\cite{Volkov2003}. The averaged volumetric map was encoded into a pdb file by SITUS vol2pdb\cite{Wriggers2010} and used for manual and automated fitting of high-resolution structural domains in UCSF Chimera and VMD\cite{Humphrey1996}. Amplitudes for individual high-resolution domains were calculated by CRYSOL\cite{Svergun1995} without fitting to experimental data to avoid artefacts from an inaccurate hydration layer. The extrapolated scattering data were used for rigid-body modelling in BUNCH\cite{Petoukhov2005} and CORAL\cite{Petoukhov2012} from the ATSAS software package. For M1.1, the NTD (residues A1--N367), the structured region S of the CTD (residues T377--E425) and the GPI anchor attached to the C-terminal residue S443 were modelled into a spatial relationship, whereas the linkers L1 (residues L368--Q376) and L2 (residues K426--G442) were computed during the rigid-body modelling as $\alpha$-carbon chains. The generated set of four models showed a good fit to the experimental data with the discrepancy $\chi^2$ ranging from 0.49--1.01. The contact conditions that were used in the computation of the initial four models varied slightly by up to 1 {\AA} and the models were created using either BUNCH or CORAL. To make sure that the conformational differences in model 2 are not a result of the computational differences, consecutive BUNCH iterations of this model (model 2a in Supplementary Fig.~4a) were calculated using identical input parameters. The resulting models, model 2b and model 2c, displayed conformations that closely resembled models 1--4 in Fig.~\ref{f:NMB3:Models}, which emphasizes the robustness of our system. Models for I1.24 were generated as described for M1.1 and the four models had a $\chi^2$ that ranged from 1.0 to 1.4.

P2 symmetry was applied to monomers of the individual domains and initial coordinates were derived from a manual fit of the domains with the respective vol2pdb map. Contact restrictions derived from the N-terminal crystal structures were applied to preserve the correct dimerization surface between the monomers in P2. Unstructured parts of the CTD were removed manually from the pdb files to take the NMR ensemble source of the C-terminal structure into account and to increase the number of blanks between structured regions to at least five residues for correct linker modelling in CORAL. A maximum scattering vector $q_\mathrm{max}$ of up to 3 nm${}^{-1}$ was chosen individually for each dataset. The GPI anchor was modelled and attached to the C-terminal amino acid using VMD and Chimera. The configuration of the S2 domain and the directly connected GPI in I1.24 were chosen to resemble the configuration of previously performed minimizations\cite{Jones2008} of these regions, whereas the GPI anchor and final C-terminal residue in M1.1 were treated as individual domains during the modelling.

\paragraph{Preparation of VSG M1.1 in supported lipid bilayers for single-molecule imaging. }
Purification, fluorescence labelling and incorporation of the membrane form VSG (mfVSG) was performed as previously described by Hartel et al.\cite{Hartel2015}. In brief, a crude mfVSG extract from the \textit{T.~brucei} wild-type strain Lister 427 was obtained using 0.1\% trifluoroacetic acid (v/v), followed by purification of the membrane form M1.1 via HPLC. The presence of the intact VSG anchor was confirmed by western blotting (10\% polyacrylamide gel, nitrocellulose membrane). The purified protein was labelled with Atto-647N NHS-ester (degree of labelling $< 0.2$). Small unilamellar vesicles (SUVs) were produced to a concentration of 1 mg ml${}^{-1}$ from rehydrated SOPC (1-stearoyl-2-oleoyl-sn-glycero-3-phosphocholine) (Avanti Polar Lipids)) in vesicle buffer (20 mM Tris-HCl, pH 7.4, 50 mM NaCl and 0.5 mM CaCl${}_2 $) by sonication or extrusion through polycarbonate membranes (50 nm pore size (Whatman)). For the analysis of the diffusion of lipids, trace amounts of DOPE-488 lipids (1,2-dioleoyl-sn-glycero-3-phosphoethanolamine) were added to the lipid mix prior to the preparation of the SUVs. Supported lipid bilayers were formed by fusion of the SUVs onto cleaned hydrophilic glass slides for 60 min at 37\degc{}. The membranes were rigorously washed with vesicle buffer to remove excess SUVs and $\sim 0.1$ pmol of labelled mfVSG M1.1 were subsequently applied onto this lipid bilayer for incorporation of the lipid moiety of the protein into the outer leaflet. Successful incorporation was confirmed by microscopy. After typically 5 min, the membrane was washed to stop the integration by removing any free protein from the buffer.

\paragraph{Single-molecule imaging.} 
An inverted widefield microscope (Leica DMI6000B) equipped with a high numerical aperture lens (HCX PL APO $100\times / 1.47$ OIL CORR TIRF) and an EMCCD camera (pixel size 16 $\mu$m (Andor Te\-chno\-logy)) was used for single-molecule imaging. Samples were illuminated with a 514 nm or 647 nm laser beam (Cobolt) for 10 ms at a mean intensity of 2 kW cm${}^{-2}$ under the control of an acousto-optic tunable filter (AOTF). The camera-ready signal triggers a function generator, which sends a rectangular signal to the AOTF, and thereby specifies the starting point and duration of the illumination. Fluorescence signals were detected by the camera using the appropriate filter combinations. Consecutive images (1,000) at the region of interest ($120\times 120$ pixels) were recorded at $T =  24$\degc{} and 28 Hz ($\Delta t = 36 $ ms). Single fluorescently labelled VSG M1.1 proteins were localized with a precision of 15 nm.

\paragraph{Analysis of single-molecule data.} 
The entire processing of data was performed in MATLAB R2014a (The MathWorks) using self-developed scripts. Single Atto-647N molecules were localized by Gaussian fitting in background-corrected images\cite{Schmidt1996} and filtered with respect to the single-molecule footprint of Atto-647N (intensity and width) and detection-error thresholds. From the remaining single-molecule positions, trajectories were computed based on a probabilistic algorithm, as described previously\cite{Hartel2015,Schmidt1996}. 
A total of $N_t=2,969$ trajectories of variable length $n_i$ ($20 \leq n_i \leq 1263$, $\langle n_i \rangle =49$) and  $N_t=1,890$ trajectories of variable length $n_i$ ($9 \leq n_i \leq 84$, $\langle n_i \rangle =13$) were extracted for proteins and lipids, respectively. Each is denoted with an index $i$, and consists of a sequence of positions $\{x^i(t_j),y^i(t_j)\}$ recorded at discrete times $t_j$. 

To analyse the traces, sequences of squared step length$(l_j^i)^2$ were constructed by calculating:
\begin{equation}
(l_j^i)^2 = \left(x^i(t_j)-x^i(t_{j-1})\right)^2 +  \left(y^i(t_j)-y^i(t_{j-1})\right)^2,
\end{equation}
in every time step for each trace. A subset of such unfiltered sequences is shown in Supplementary Fig.~8a. 

\paragraph{}\noindent\emph{Filtering non-Gaussian behaviour.} 
The strong separation in the characteristic step length is used as a main criterion to filter this type of motion. Furthermore, as transitions from and into this slowest behaviour are commonly observed, the following procedure was adopted. First, the mean step length $\langle L^2 \rangle$  was calculated over the entire set of data:
\begin{equation}\label{eq:lsquared}
\langle L^2 \rangle = \frac{1}{\sum_{i=1}^{N_\mathrm{t}} n_i} \sum_{i=1}^{N_\mathrm{t}} \sum_{j=1}^{n_i} (l_j^i)^2.
\end{equation}

In the second step, a threshold $T$ was set to $T = 0.15 \langle L^2 \rangle $, and, from the whole set, all sequences of at least ten successive steps with a square length smaller than $T$ were extracted. As a result of such filtering, sequences were obtained that were either segments of the initial traces or whole trajectories. In the next step, a new $N_\mathrm{t}$ was calculated together with its respective $\langle L^2 \rangle $, and the extraction was repeated. The procedure was reiterated until the value of $\langle L^2 \rangle $ saturated (to the value of 0.06 \musq{} for the current data). The obtained sequences form a set (black) with a purely exponential distribution of displacements (Supplementary Fig.~9c). Owing to its non-Gaussian nature, this subclass of trajectories was removed from the current analysis. The remaining set (exemplary traces shown in Fig.~\ref{f:NMB5:DiffusionBehaviour}a) is characterized by $\langle L^2 \rangle = \langle L_{\mathrm{Gauss}}^2 \rangle $. 

\paragraph{}\noindent\emph{Separating the slow and fast diffusion of proteins.} Given that the two remaining populations, denoted as slow and fast (brown and green, respectively, in Fig.~\ref{f:NMB5:DiffusionBehaviour}a), have significant overlap, the separation of traces is more delicate. Therefore, besides thresholding, an additional statistical criterion, based on the calculation of probability likelihoods, was introduced in the analysis. Furthermore, as the transitions between the two populations were not readily observed in this set of data (most trajectories were significantly shorter than the transition time between the conformations), the analysis was performed on the entire traces. 

Calculation of likelihoods relies on determining the probability density $P(l^2)$ for observing a particular square step length $l^2$, and then finding the probability for a given sequence of step lengths to belong to that distribution---for example, for each sequence $\{ (l^i)^2 \} = \{ (l_1^i)^2,\dots,(l_{n_i}^i)^2  \} $, the likelihood is given by the product  of probability densities $P((l_j^i)^2)$ to find all the elements of the sequence in a given distribution. Calculating the ratio of likelihoods to belong to hypothetical distributions  $P_{\mathrm{A}}$ and  $P_{\mathrm{B}}$:
\begin{equation}
R_i(P_\mathrm{A},P_\mathrm{B})=\frac{\mathcal{L} \left(P_\mathrm{A}|\{ (l^i)^2 \} \right)}{\mathcal{L} \left(P_\mathrm{B}|\{ (l^i)^2 \} \right)} 
= \prod_{j=1}^{n_i} \frac{P_\mathrm{A} ((l_j^i)^2)}{P_{\mathrm{B}}((l_j^i)^2)},
\end{equation}
provides a unique criterion to separate two populations. Namely, $R_i=1$ signifies that the probability to belong to both distributions is the same. On the other hand, $R_i>1$ means that it is $R_i$ times more likely to belong to the distribution  $P_{\mathrm{A}}$. In an analogous way, $R_i < 1$ suggests that it is $1/R_i$ more likely that a sequence belongs to $P_{\mathrm{B}}$ than to $P_{\mathrm{A}}$.

To separate the slow and fast diffusion, in the first step each trace with a mean step length $\langle (l^i)^2 \rangle  < \frac{2}{3} \langle L_{\mathrm{Gauss}}^2 \rangle = 0.19$ $\mu$m$^2$ was considered to belong to the population of slowly diffusing proteins. The remaining sequences were used as the initial set associated with proteins that exhibited fast diffusion. In the second step, distributions of square step lengths $P_{\mathrm{fast}}({l}^2)$ and $P_{\mathrm{slow}}({l}^2)$ were constructed. Both distributions showed a simple exponential behaviour, as expected for simple Gaussian diffusion, with the fast process having a fourfold larger decay length. The ratio of likelihoods  $R_i$  was extracted for each trace $i$, and the traces were reassigned to the appropriate family (with $R_i(P_{\mathrm{fast}},P_{\mathrm{slow}}) =1$  as the threshold). In the next step, the probability distributions $P_{\mathrm{fast}}$ and $P_{\mathrm{slow}}$ were re-evaluated and $R_i$ values recalculated. This step was reiterated until no further swaps between families occurred. The converged, purely exponential distributions are shown in Supplementary Fig.~8b, and the Gaussian distributions of 1D displacements are shown in Supplementary Fig.~9a,b.

\paragraph{}\noindent\emph{Comparison with the likelihood that a unique distribution describes both fast and slow diffusion of proteins.} The described procedure allowed the characterization of each trace $i$ by the set $ \{\langle (l^i)^2 \rangle, R_i\}$, as shown in Fig.~\ref{f:NMB5:DiffusionBehaviour}b, where the slow and fast processes are clearly separated into two clusters. Importantly, the same clustering would take place if the data were sorted by $R_i(P_{\mathrm{total}},P_{\mathrm{slow}}) =1$ as the threshold, where $P_{\mathrm{total}}$ signifies an exponential fit to the entire ensemble of step lengths. This last finding was consistent with the observation of two distinct subclasses of trajectories. 

\paragraph{}\noindent\emph{Calculation of the mean square displacements.} To generate the distribution of diffusion coefficients $D_i$ for each trace, the mean square displacement MSD${}_i = \langle r_i^2 (t) \rangle$ was calculated as a function of time:
\begin{align}
\textrm{MSD}_i &= \frac{1}{n_i-t/\Delta t}  \sum_{j=1}^{n_i-t/\Delta t} \left[  \left( x^i (t_j+t) -x^i(t_j) \right)^2 \right. \nonumber\\ 
& \hspace{3.5cm} \left. +  \left( y^i (t_j+t) -y^i(t_j) \right)^2 \right]  \\
&= 4 D_i t.
\end{align} 
Here $t$ is a discrete time defined as  $t=N\Delta t$, where $N\in\mathbb{N}$. Owing to the limited size of each trajectory only the first three points were considered statistically relevant and used for linear regression, with  $D_i$  being a fit parameter. The distribution of  $D_i$  is shown for both families of traces in Fig.~\ref{f:NMB5:DiffusionBehaviour}c.

A more-detailed analysis of the properties of the distribution allowed the calculation of the mean $\langle D \rangle$, standard deviation $ \sigma_D$ and standard error $\mathrm{SE}_D$:
\begin{equation}
\langle D \rangle = \frac{1}{N_\mathrm{t}} \sum_{i=1}^{N_\mathrm{t}} D_i, 
\end{equation}
\begin{align}
\sigma_D &= \sqrt{\frac{1}{N_\mathrm{t}} \sum_{i=1}^{N_{\mathrm{t}}} (D_i- \langle D \rangle )^2 }, \\
 \mathrm{SE}_D &=\sqrt{\frac{\sigma_D }{N_\mathrm{t}}} .
\end{align}
Analogous results (shown in the inset of Fig.~\ref{f:NMB5:DiffusionBehaviour}c) were obtained by calculating the time-averaged MSD on the cumulative trajectory for each population. The latter was constructed by concatenating all the trajectories that belonged to a population and creating a single long trace of length $N_\mathrm{c}$. In this case, the MSD was calculated as
\begin{align}
\textrm{MSD} &=  \frac{1}{N_c-t/\Delta t}  \sum_{j=1}^{N_c-t/\Delta t} \left[  \left( x(t_j+t) -x(t_j) \right)^2 \right. \nonumber \\
& \hspace{3.5cm} \left. + \left( y(t_j+t) -y(t_j) \right)^2 \right] \\
& = 4 D t.
\end{align}

\paragraph{Data availability.} Structure information for the NTD and CTD of VSG M1.1 have been deposited in the RCSB Protein Data Bank with the accession codes PDB \href{https://www.rcsb.org/structure/5LY9}{5LY9} and PDB \href{https://www.rcsb.org/structure/5M4T}{5M4T}, respectively. Other data that support the findings of this study are available on request.

\section{Acknowledgements}
This work was supported by the Deutsche Forschungsgemeinschaft (DFG, grants EN 305, GRK 1114 to M.~E.~and SFB 630 to M.~E.~and C.~K.), and the Wellcome Trust (grant 022758/Z/03/Z to M.~Ca.). A.-S.~S.~and M.~Cv.~were funded from grant ERC StG 2013--337283 of the European Research Council and supported by the DFG GRK 1962. M.~E.~is a member of the Wilhelm Conrad R{\"o}tgen-Center for Complex Material Systems. We thank the Helmholtz-Zentrum Berlin for the allocation of synchrotron radiation beamtime and the staff of the BESSY at beamline 14.1 for technical assistance. The SAXS experiments were performed on beamline BM29 at ESRF. We are grateful to A.~Round at the ESRF for providing assistance in using beamline BM29 and for invaluable tips concerning data analysis. We thank D.~Nietlispach for the acquisition of NMR data and B.~Morriswood for critical reading of the manuscript.

\section{Author contributions}
T.~B., N.~G.~J.~and M.~E.~conceived the study, T.~B., N.~G.~J., M.~Ca., A.-S.~S., S.~F., C.~K.~and M.~E.~designed the research; T.~B., N.~G.~J., M.~G.~and S.~F.~performed the experiments; T.~B., N.~G.~J., C.~S., M.~Cv., M.~G., H.~R.~M., J.~K., M.~B., A.-S.~S., S.~F.~and M.~E.~analysed the data; T.~B., N.~G.~J.~and M.~E.~wrote the paper with contributions from M.~Cv., A.-S.~S.~and S.~F.~during manuscript editing.

\section{Additional information}
Supplementary information and supplementary video is available for this paper at doi: \href{http://dx.doi.org/10.1038/s41564-017-0013-6}{10.1038/s41564-017-0013-6}.

This is the accepted manuscript of the paper published in \emph{Nature Microbiology} \textbf{2}, pp.~1523--1532 (2017).

\bibliographystyle{myunsrtnat}
\bibliography{NatureBibliography}

\end{document}